\documentclass[12pt]{article}

\advance \textheight by \topskip
\makeatletter
\@addtoreset{equation}{section}
   
\makeatother

\def\vx{\vec{x}}
\def\vy{\vec{y}}
\def\vX{\vec{X}}
\def\vp{\vec{p}}
\def\vQ{\vec{Q}}

\def\vJ{\vec{J}}

\usepackage{amsmath,amssymb}
\begin{document}

\title{
\begin{flushright}
{\small USACH-FM-02/01}\\[1.0cm]
\end{flushright}
{\bf   Non-relativistic anyons,
exotic Galilean symmetry and noncommutative plane
}}

\author{{\sf Peter A. Horv\'athy}\thanks{
Permanent address: {\it Laboratoire de Math\'ematiques et de
Physique Th\'eorique}, Universit\'e de Tours (France).
E-mail: horvathy@univ-tours.fr.}
{\sf\ and Mikhail S. Plyushchay}\thanks{
Also: {\it Institute for High Energy Physics},
Protvino (Russia).
E-mail: mplyushc@lauca.usach.cl}
\\[4pt]
{\small \it
Departamento de F\'{\i}sica,
Universidad de Santiago de Chile}\\
{\small \it Casilla 307, Santiago 2, Chile}\\
}
\date{}

\maketitle

\begin{abstract}
We show that the Lukierski et al.
model, invariant with respect to the
two-fold centrally
extended Galilei group,
can be decomposed into an
infinite number of independent copies
(differing in their spin) of
the ``exotic'' particle of Duval et al.
The difference between the two models is
found to be
sensitive to electromagnetic coupling.
The nature of
the noncommutative plane
coordinates is discussed in the light of the exotic Galilean
symmetry. We prove that the first model,
interpreted as describing
a non-relativistic anyon,
is the non-relativistic
limit of a particle with torsion related
to relativistic anyons.

\end{abstract}

\vskip.5cm

As it has been known for some time,
the planar Galilei group admits an ``exotic'' two-parametric
central extension. Recently, two classical systems have been
presented that exhibit this extended Galilean symmetry.
One of them, put forward by Lukierski, Stichel and
Zakrzewski \cite{LSZ}, uses
the second-order Lagrangian
\begin{equation}
L_{LSZ}=\frac{1}{2}m\dot{\vx}^2
+
\frac{\kappa}{2}\varepsilon_{ij}\dot{x}_i\ddot{x}_j,
\label{LSZlag}
\end{equation}
where $m$, the mass, and $\kappa$, the ``exotic" parameter,
label the central extension \cite{doublegal}. This system
requires
a $6$-dimensional phase space.

The other model \cite{Grig, Duval, DH}
is derived from the ``exotic'' Galilei group following
Souriau
\cite{SSD}, who identifies classical ``elementary'' systems
with
the coadjoint orbits of the group, endowed with their
canonical
symplectic structures. These orbits are $4$-dimensional, and
depend on $4$
parameters denoted by $s$, $h_0$, $m$ and $\kappa$. Their
symplectic structure
induces on the $5$-dimensional ``evolution space'' made of
positions, momenta, and
time \cite{SSD} the closed $2$-form
\begin{equation}
\omega
   =d{\vp}\wedge d\vx
+
\frac{\kappa}{2m^2}\varepsilon_{ij}dp_i\wedge dp_j
-
\frac{\vp\cdot d\vp}{m}\wedge dt.
\label{exosymp}
\end{equation}
This (free) model is equivalent  to that
of a  particle in the non-commutative plane \cite{
NC,noncomgeom,Moy}
when $\kappa/m^2$ is identified
(as  always in what follows)
with the ``non-commutative parameter'' $\theta$ \cite{DH}.
Coupled to a gauge field,
the ``exotic'' model
(\ref{exosymp})
has been used to derive the ground
states
of the Fractional Quantum Hall Effect \cite{DH}.
\goodbreak

The two-form (\ref{exosymp}) depends only on
$m$ and $\kappa$; the two other parameters show up
only in the
way the Galilei group acts on the orbit,
i.e.,
in the associated conserved quantities. These latter are
found
to be the momentum, $\vp$,\goodbreak
\begin{equation}
\begin{array}{cc}
{\cal J}=\vx\times\vp+\displaystyle\frac{\theta}{2}\vp{}^2+
s,\hfill
&\hbox{angular momentum,}\hfill
\\[6pt]
{\cal K}_i=mx_i-p_it+m\theta\varepsilon_{ij}p_j,\qquad\hfill
&\hbox{boosts,}\hfill
\\[4pt]
H=\displaystyle\frac{\vp^2}{2m}+h_0,\hfill
&\hbox{energy,}\hfill
\end{array}
\label{EXconsquant}
\end{equation}
supplemented with
$m$ and $\kappa=m^2\theta$.
The parameter
$s$ is hence interpreted as {\it anyonic spin}, and
$h_0$ is the {\it internal energy}. In \cite{DH}
$s=h_0=0$.
\goodbreak

The two models look rather different.
Below we show, however, that the model  (\ref{LSZlag})
is in fact composed of an infinite number of independent
copies (differing in their spin) of the model
(\ref{exosymp}), (\ref{EXconsquant}),
and point out that the difference between
the models
is sensitive to  electromagnetic coupling.
We discuss the nature of the
noncommutative plane coordinates
in the light of the exotic Galilean
symmetry, and argue that the model (\ref{LSZlag})
represents {\it nonrelativistic anyons}.
We demonstrate also that the model (\ref{LSZlag})
is in fact the non-relativistic limit
of the particle with torsion \cite{MPtor},
which underlies the relativistic anyons.
\vskip 0.5cm

Let us start with introducing
Lagrange multipliers $p_i$
and new variables $y_i$;
then, adding the term
$p_i(\dot{x}_i-y_i)$ to (\ref{LSZlag}) yields
the
first-order Lagrangian
\begin{equation}
L_I=
\vp\cdot\dot{\vx}+\frac{\kappa}{2}\vy\times\dot{\vy}
+
\frac{m\vy^2}{2}-\vp\cdot\vy
\label{FaJalag}
\end{equation}
   equivalent to (\ref{LSZlag}). Introducing the coordinates
\begin{equation}
X_i=x_i+
m\theta\varepsilon_{ij}y_j
-
\theta\varepsilon_{ij}p_j,
\qquad
Q_i=
   \theta(my_i-p_i)
\label{LSZgoodcoordinates}
\end{equation}
and $p_i$ on the
$6$-dimensional phase space
allows us to present
the symplectic structure and the
Hamiltonian associated with
(\ref{LSZlag}) through
(\ref{FaJalag}) as
\begin{equation}
\begin{array}{ll}
\Omega=dp_i\wedge dX_i
+
\displaystyle\frac{\theta}{2}\varepsilon_{ij}dp_i\wedge dp_j
+
\displaystyle\frac{1}{2\theta}\varepsilon_{ij}dQ_i\wedge
dQ_j,
\\[6pt]
H=\displaystyle\frac{\vp^2}{2m}-\displaystyle\frac{1}{2m
\theta^2}\vQ^2.
\end{array}
\label{LSZsympham}
\end{equation}

Our coordinates $\vX$
are similar to but still different from those, $\vX^{L}$, of
Lukierski et al.;
in fact $X_i=X^{L}_i-\theta\varepsilon_{ij}p_j$.
Note that the ``internal'' coordinates, $\vQ$,
   measure  the amount by which  the ``external'' momentum,
$\vp$,
differs from ($m$-times) the velocity, $\vy=\dot{\vx}$.
In the coordinates
(\ref{LSZgoodcoordinates}),
the ``external'' and ``internal'' motions,
described by
$\vX$ and $\vQ$, respectively, are decoupled:
\begin{equation}
\dot{X}_i=\frac{p_i}{m},
\qquad
\dot{p}_i=0,\qquad
\dot{Q}_i=\frac{1}{m\theta}\,\varepsilon_{ij}\,Q_j.
\label{LSZeqnmotion}
\end{equation}
Thus, while the external motion is free, $\ddot{\vX}=0$,
the internal coordinates perform a uniform rotation with
angular velocity
$(m\theta)^{-1}$. By (\ref{LSZgoodcoordinates}), the
original coordinates
$x_i=X_i-\varepsilon_{ij}Q_j$
rotate hence, in general, around the uniformly translating
guiding center,
cf. \cite{LSZ}\footnote{
It is interesting to observe the
analogy with the
motion of a usual spinless charged particle
in  combined constant magnetic and
electric fields, though the particle here is free.}.
   Such a ``Zitterbewegung'' would be most
surprising for a free non-relativistic particle, and we
argue that it is the new coordinate $\vX$, and  not the
original ones, $\vx$, that should be viewed as
physical.
Note, however, that while original coordinates
commute, $\big\{x_i,x_j\big\}=0$,
both the ``external'' and the ``internal'' positions,
$\vX$ and $\vQ$, respectively, are non-commuting,
\begin{equation}
\big\{X_i,X_j\big\}=\theta\varepsilon_{ij},
\qquad
\big\{Q_i,Q_j\big\}=-\theta\,\varepsilon_{ij}.
\label{commrel}
\end{equation}

The action of the Galilei group on $\vX$ and $\vp$,
deduced from the natural action using
(\ref{LSZgoodcoordinates}),
is  conventional.
   The internal coordinate $\vQ$ is left invariant in turn
   by
the Galilei boosts, the rotations act
on it as on a vector, $\vQ\to R\vQ$.
The system is invariant with respect to this action,
allowing us to
recover
the conserved quantities (equivalent to those in
   \cite{LSZ}), namely $\vp$, augmented with
\begin{equation}
\begin{array}{ll}
{\cal J}=\vX\times\vp+\displaystyle\frac{\theta}{2}\vp^2
+
\displaystyle\frac{1}{2\theta}\vQ^2,\hfill
\\[10pt]
{\cal K}_i=mX_i-p_it+m\theta\varepsilon_{ij}p_j,\qquad\hfill
\\
\end{array}
\label{LSZconsquant}
\end{equation}
and the energy, $H$, in (\ref{LSZsympham}).
This latter can also be represented
as
\begin{equation}
H=\frac{2}{m}\vp{}^2-\frac{1}{m\theta}
({\cal J}-m^{-1}\varepsilon_{ij}{\cal K}_ip_j).
\label{HJPK}
\end{equation}
This means that like in the case of the
usual free non-relativistic spinless
particle ($\kappa=0$), the energy is
defined in terms of other
generators of the Galilei group.
Let us also observe that
the coordinates $X_i$
form a {\it Galilean vector} since
they have the correct transformation property
under boosts, namely
$
\big\{{\cal K}_i,X_j\big\}=-\delta_{ij}t.
$

For comparison with \cite{LSZ}, we mention that
though in the coordinates $X^L_i$
the boosts appear in  a more simple form,
${\cal K}_i=mX^L_i-p_it$, their brackets
are different from (\ref{commrel}) in sign,
$\{X^L_i,X^L_j\}=-\theta\varepsilon_{ij}$,
and unlike our $X_i$, they satisfy the relation
$
\big\{
{\cal K}_i,X_j^L\big\}=-\delta_{ij}t-m\theta\varepsilon_{ij}
$,
which means that they {\it do not form} a Galilean vector.

Our clue is to observe that the system
(\ref{LSZsympham})
can consistently be
restricted to
the surface
\begin{equation}
S_C:
\qquad
\vQ^2-C^2=0,\quad C={\rm const}.
\label{constraintsurf}
\end{equation}
For $C=0$, in particular, $S_0$ is a $4$-{\it dimensional}
surface: the
internal motion is reduced to a single point. Then the
restriction of
(\ref{LSZsympham})
yields the symplectic form
$
\Omega_0=dp_i\wedge dX_i
+
(\theta/{2})\varepsilon_{ij}dp_i\wedge dp_j
$
and the Hamiltonian
$H_0=\vp^2/2m$
respectively, so that $\Omega_0-dH_0\wedge dt$
is  the ``exotic'' $2$-form (\ref{exosymp}).
The  coordinates $\vX$ are reduced now in fact to
the original ones that
perform  the usual free motion with {\it no whirling}
around, and
can, therefore, be viewed as physical.
The conserved quantities (\ref{LSZconsquant}) become,
furthermore, precisely those found in
\cite{DH}, i.e. (\ref{EXconsquant}) with no anyonic spin
and vanishing
internal energy, $s=0$ and $h_0=0$, respectively.
Note also that, consistently with (\ref{commrel}),
the original coordinates
$x_i$ become also non-commuting, owing to the
constraint (\ref{constraintsurf}).
This result is understood by observing that
the action of the Galilei group on $6$-dimensional phase
space leaves the surfaces $S_C$ invariant.
For $C=0$ it is transitive as well as symplectic.
By Souriau's theorem \cite{SSD}
$S_0$, endowed with the restricted two-form $\Omega_0$, is
hence
equivariantly symplectomorphic to a coadjoint orbit of the
group,
namely to $\omega$ in (\ref{exosymp}) and
(\ref{EXconsquant}), with $s=h_0=0$.
Note also that when $\vQ^2=0$, the momentum is $m$ times the
velocity, $\vp=m\dot{\vx}$. The second-order term in
(\ref{LSZlag}) becomes  the ``exotic'' term
$(\theta/2)\varepsilon_{ij}p_i\dot{p}_j$,
and the Lagrangian (\ref{FaJalag}) reduces to
the one used in \cite{DH}.

Let us now turn to the $5$-dimensional surfaces $S_C$ with
$C\neq0$.
The coordinates $\vX$ move still freely but
the original coordinates, $\vx$, suffer the Zitterbewegung
found above.
An interesting insight can be gained by studying
in the variational aspects of the system.
   The  two-form and Hamiltonian in (\ref{LSZsympham})
correspond indeed to the variational
(Cartan \cite{SSD}) $1$-form (consistent with
(\ref{FaJalag})),
\begin{equation}
\Gamma=
\vp\cdot d\vX+\frac{\theta}{2}\vp\times d\vp
+
\frac{1}{2\theta}\vQ\times d\vQ
-
\left(\frac{\vp^2}{2m}
-
\frac{1}{2m\theta^2}\vQ^2\right)dt.
\label{cartanform}
\end{equation}
Its restriction to $S_C$ is plainly
\begin{equation}
\gamma=
\vp\cdot d\vX+\frac{\theta}{2}\vp\times d\vp
+
\frac{C^2}{2\theta} d\sigma
-
\left(\frac{\vp^2}{2m}
-
\frac{1}{2m\theta^2}C^2\right)dt,
\label{redcartanform}
\end{equation}
where we have parametrized ${\cal Q}=Q_1+
iQ_2$ as ${\cal Q}=Ce^{i\sigma}$.
Note that there is no equation of motion for the residual
internal variable $\sigma$ which became now unphysical.

The action of the Galilei group is consistent with the
constraint,
so it is still a symmetry, yielding once again the conserved
quantities (\ref{LSZconsquant})
with $\vQ^2$ replaced by the constant $C^2$.
We obtain hence the shifted moment
map \cite{SSD} (\ref{EXconsquant})
with anyonic spin $s=C^2/2\theta$ and internal energy
$h_0=-C^2/2\theta^2$, whose presence comes from that, when
our
$\vX$ is identified with the physical coordinate in
(\ref{exosymp}),
then (\ref{redcartanform}) only differs from
the variational $1$-form used in \cite{DH} in the (exact)
``internal'' $1$-form $(C^2/{2\theta})d\sigma$ and in a
shift in the Hamiltonian.
   Note that, curiously, the internal degrees of freedom,
   $\vQ$,
contribute a {\it negative} term to the energy.

Physically, the  restriction to $S_C$ yields
non-relativistic
{\it anyons}.

The  exterior derivative
of the variational $1$-form (\ref{redcartanform}) to the odd
dimensional
manifold $S_C$ is singular;
its kernel corresponds to $\sigma$.
Factoring this out,
   the residual $4D$ space can be parametrized by
$\vX$ and $\vp$, upon which the (extended) Galilei group
acts transitively,
so that Souriau's theorem \cite{SSD} identifies it, once
again, with a coadjoint orbit with
   symplectic structure (\ref{exosymp}).

Note that the spin and the internal energy derived above are
linked, $h_0=-(m\theta)^{-1}s$. It is rather easy to
modify the original model \cite{LSZ} to make them
independent:
it is enough to add to $L_{LSZ}$ the spin term
\begin{equation}
L_s=
s_0\varepsilon_{ij}\frac{\dot{x}_i\ddot{x}_j}{\dot{\vx}^2}.
\label{spinterm}
\end{equation}
This term is in fact $s_0d\varphi$, where $\varphi$ is the
polar angle of
the planar velocity vector $\dot{\vx}$.
This ``Aharonov-Bohm-type'' term is topological: it
merely changes the action
by $2\pi s_{0}\times $(winding number),
where the winding number labels the homotopy class
i.e.
counts the number of times the velocity space
curve winds around the
origin.
It does not change
the equations of motion (although it contributes
at the quantum level).
Taking into account the way the Galilei group
acts, is simply results in adding $s_0$ to the conserved
angular
momentum, yielding the  general case in
(\ref{EXconsquant}). Note also that the exact ``internal''
term
$(C^2/2\theta)d\sigma$ in (\ref{redcartanform}) is actually
also of a similar form: it is indeed
$(C^2/2\theta){\varepsilon_{ij}Q_i\dot{Q}_j}/{\vQ^2}$.

Let us mention that from
the $1$-form (\ref{cartanform}) and the
relation (\ref{constraintsurf}),
   the technique of Dirac constraints allows us to
derive the Lagrangian
\begin{equation}
L_C=
m\vy\cdot\dot{\vx}-\frac{m\vy^2}{2}
+\frac{\kappa}{2}\vy\times\dot{\vy}
-mC\vert\dot{\vx}-\vy\vert,
\label{constrFaJalag}
\end{equation}
for which
(\ref{constraintsurf})
appears as a first class constraint.
The latter generates gauge
transformations
$\sigma\rightarrow \sigma+\alpha(t)$
which mean that the circular coordinate $\sigma$
is a pure gauge variable. As a result,
Lagrangian (\ref{constrFaJalag})
corresponds to the reduced system.
Note that for any value of $C$,
the coordinates $X_i$, unlike $x_i$,
are the  gauge-invariant variables (commuting
with the first class constraint
(\ref{constraintsurf})),
and that for $C=0$, (\ref{constrFaJalag})
reduces  to the Lagrangian
of the model (\ref{exosymp})
under the identification
$\vp=m\vy$, cf. (\ref{redcartanform}).

By virtue of the decomposition into external and internal
spaces, the quantization of the model is straightforward.
The external space is conveniently quantized using
the {\it commuting} position-like variables
\begin{equation}
\tilde{X}_i=X_i+\frac{\theta}{2}\varepsilon_{ij}p_j,
\label{commcoord}
\end{equation}
yielding the quantized operators at once
\cite{DH}.
Let us stress, however, that like the coordinates
$X^L_i$,
the commuting variables
$\tilde{X}_i$ are {\it not} physical, since they transform
incorrectly under Galilean boosts, namely as
$
\big\{{\cal K}_i,\tilde{X}_j\big\}=
-\delta_{ij}t-(m\theta/{2})\varepsilon_{ij}.
$
So, the coordinates $X_i$ describing the noncommutative
plane are  identified uniquely as the coordinates
forming a Galilean vector.

The  internal space is the symplectic plane; its
quantization is hence conveniently achieved using
the `` Bargmann-Fock'' framework, which yields the internal
wave functions $f({\cal Q})e^{-\vert {\cal Q}\vert^2/4}$
where $f({\cal Q})$ is holomorphic.
The internal Hamiltonian,
$H_{int}=-\vQ^2/2m\theta$,
can be viewed as the Hamiltonian of a $1$-dimensional
oscillator
with {\it phase space} coordinate
${\cal Q}$, {\it negative}
mass $-m$ and frequency
$\vert m\theta\vert^{-1}$. Its spectrum is, therefore,
$E_{int}=-(m\vert\theta\vert)^{-1}(n+1/2)$,
cf. \cite{LSZ}.
The divergent negative
energies can be eliminated by constraining the system to
$S_C$,
which only leaves us with a constant (negative) shift;
requiring
positivity \cite{LSZ} amounts to setting $C=0$.

Since only the rotations act effectively on the internal
space,
only
this latter contribute, namely adding
   $s=-m\theta E_{int}$ to the conserved angular momentum.
When restricted to $S_C$, the radial oscillations disappear
and $s$ becomes the (constant)
anyonic spin. Let us
stress that $s=C^2/2\theta$ can take
any real value.

When we put the system into an electromagnetic field,
the result crucially depends on the way the coupling is
defined \cite{DH}.
Lukierski et al. only consider the coupling to a scalar
potential.
Their rather strange-looking, velocity-dependent
   expression is chosen so that, when expressed in their ``
   good''
coordinates $\vX^L$, it becomes simply $V(\vX^L)$.
Here we posit instead our coupling so that it becomes
natural
when expressed in our coordinates $\vX$.
Then, in the case of the scalar potential coupling,
the quantization rule
(\ref{commcoord})
results in the Moyal product term in the Schr\"odinger
equation,
\begin{equation}
V(X)\Psi(\tilde{X})=V\left(\tilde{X}_i-\frac{\theta}{2}
\varepsilon_{ij}p_j\right)\Psi(\tilde{X})=V(\tilde{X})\star
\Psi(\tilde{X}),
\label{Moyal}
\end{equation}
that reproduces correctly the structure of the
one-particle sector of the quantum theory
on the noncommutative plane \cite{Moy,noncomgeom,DJ}.

Now, there is no difficulty to include a magnetic
field: we simply
apply Souriau's prescription \cite{SSD} who suggests to add
to the
exterior derivative of the free variational form
(\ref{cartanform}) the electromagnetic two-form
$F=(1/2)F_{\mu\nu}(X)dX^\mu\wedge dX^\nu$,
$X^\mu=(t,\vX)$,
expressed in the ``natural'' position coordinates. This
coupling is plainly
gauge invariant and, (unlike for the
rule proposed in \cite{NC}), the Jacobi identities hold for
non-constant fields,
provided $F$ merely satisfies the homogeneous Maxwell
equations, $dF=0$.
Applying this rule yields unchanged internal motion; as to
the external motion, we recover
the model studied in detail in \cite{DH},
which, in the critical case $eB\theta=1$, yields the ground
states of the Fractional Quantum Hall Effect.
Note also that the coupling
of relativistic anyons to gauge fields have been considered
in
\cite{CNP}.

Here, we confine ourselves
with presenting how this coupling rule
--- natural in our framework ---
appears in the context of the model \cite{LSZ}. In terms of
the
coordinates
$x_i, y_i$ and $p_i$ used in (\ref{FaJalag}) we get,
for a constant magnetic  field $B$ (in the symmetric gauge),
   the rather complicated-looking phase-space Lagrangian
\begin{equation}
\begin{array}{ll}
L_B=
p_i\left(\displaystyle\frac{m^*}{m\ }\dot{x}_i-y_i\right)
+
\displaystyle\frac{m\vy^2}{2}
+
\displaystyle\frac{m^2\theta}{2}\varepsilon_{ij}y_i\dot{y}_j
\\[8pt]
-
\displaystyle\frac{e\theta^2B}{m}
\varepsilon_{ij}p_i\dot{y}_j
+
\displaystyle\frac{eB}{2}\varepsilon_{ij}
\big(x_i+m\theta\varepsilon_{ik}y_k\big)
\big(\dot{x}_j+m\theta\varepsilon_{il}\dot{y}_l\big)
+
\displaystyle\frac{e\theta^2}{2}B\varepsilon_{ij}p_i\dot{p}_
j,
\end{array}
\label{horror}
\end{equation}
where $m^*=m(1-eB\theta)$ is the effective
mass term
introduced
in \cite{DH}.
This formula, hopeless as it is, reveals nonetheless
at least two essential points.
Firstly, putting  the effective mass to vanish, $m^*=0$,
switches off the kinetic term;
the system becomes singular, necessitating the reduction
procedure described in \cite{DH}. Secondly,
the appearance of $\dot{p}_i$ in the last term implies that
the momentum {\it ceases to be a Lagrange multiplier} and
becomes rather fully dynamical. Therefore, it can not come
by
Legendre transformation
from any Lagrangian expressed in terms of
$\vx$ and its derivatives alone.
\goodbreak

\goodbreak
Other consistent couplings are also possible, though.
One could add naively, for example, the standard constant
magnetic term
to the original Lagrangian, i. e, replace $L_{LSZ}$  by
\begin{equation}
\tilde{L}_B=
\frac{1}{2}m\dot{\vx}^2
+
\frac{\kappa}{2}\varepsilon_{ij}\dot{x}_i\ddot{x}_j
+
\frac{eB}{2}\varepsilon_{ij}x_i\dot{x}_j.
\label{LSZBlag}
\end{equation}
This would yield three types of motions, depending on the
sign of
\begin{equation}
\mu^2=1-4eB\theta.
\label{newparameter}
\end{equation}
For $\mu^2>0$ we get in fact a superposition of uniform
circular motions with frequencies
$\omega_{\pm}=-(2m\theta)^{-1}(1\pm\mu)$.
For $\mu^2<0$, $z=x_1+ix_2$ moves according to some
superposition of
$\exp[-i(2m\theta)^{-1}t]\exp[\pm\vert\mu(2m\theta)^{-1}
\vert t]z_0^{\pm}$,
which are rotational
motions with increasing or decreasing radii.
For $\mu^2=0$ we get finally a superposition of a uniform,
constant-radius
rotation, $\exp[-i(2m\theta)^{-1}t]z_0$, with a
rotation
$t\exp[-i(2m\theta)^{-1}t]\tilde{z}_0$ whose radius is
linear in time.
None of the motions with $\mu^2\leq0$ appears to be
physical;
such a coupling has therefore a limited interest.
Note that the critical value $\mu=0$ has appeared before in
a related but slightly different context \cite{Gamboa}.

Since the higher-derivative model (\ref{LSZlag}) has been
interpreted
as describing non-relativistic anyons,
we arrive at the natural question of
its relation to the higher derivative
model of the particle with torsion \cite{MPtor},
which underlies relativistic anyons and whose
action reads
\begin{equation}
{\cal A}=\int(-m+\alpha \varrho)ds,
\qquad
\varrho=\frac{\epsilon^{\mu\nu\lambda}x'_\mu
x''_\nu x'''_\lambda}{x''{}^2},
\label{Ptor}
\end{equation}
where the prime means derivation with respect to arc length,
$x'_\mu=dx_\mu/ds$,
$ds^2=-dx^\mu dx^\nu\eta_{\mu\nu}$,
$\eta_{\mu\nu}= diag (-1,1,1)$, and
$\epsilon^{0ij}=\varepsilon_{ij}$.

The action
(\ref{Ptor}) with
$\alpha=1/2$
appeared originally
in a Euclidean version
in the context of the Bose-Fermi transmutation mechanism
\cite{Pol}.
Like the model (\ref{LSZlag}), the system (\ref{Ptor})
possesses dynamical spin degrees of freedom,
and reveals the Zitterbewegung.
In analogy  with the possible energy values
$E>0$, $E=0$ and $E<0$
of the Hamiltonian (\ref{LSZsympham}),
the model (\ref{Ptor}) possesses
massive,  massless,
and tachyonic
solutions, and its classical motions
are  similar to those of the
$(2+1)$D relativistic charged scalar particle moving
in  constant electric and magnetic
fields  \cite{MPtorelm}.
Finally, it was found that the
reduction of the model (\ref{Ptor})
to the surface of constant spin
results in a relativistic anyon model
\cite{MPtor,MPany, MPJN}.
Such a  formal analogy is, of course,
not accidental, and below we
prove that the model (\ref{LSZlag})
is the non-relativistic limit of the relativistic
model (\ref{Ptor}).

In an arbitrary parametrization
$x_\mu=x_\mu(\tau)$,
the structure of the equations of motion
for the system (\ref{Ptor}),
\begin{equation}
\dot p_\mu=0,\quad
p_\mu=\displaystyle \frac{m}{\alpha}J_\mu -
\displaystyle \frac{1}{\alpha q}\epsilon_{\mu\nu\lambda}
J^\nu\dot{J}{}^\lambda,
\quad
J_\mu=\displaystyle\frac{\alpha}{q}\dot{x}_\mu,\quad
q=\sqrt{-\dot{x}{}^2},
\label{RPTeq}
\end{equation}
is similar to that
of the equations (\ref{LSZeqnmotion})
written in the initial coordinates $\vx$,
\begin{equation}
\dot{\vp}=0,\quad
p_i=my_i+
\theta m^2\varepsilon_{ij}
\dot y_j,\quad
\vy=\dot{\vx}.
\label{pxy}
\end{equation}
Note that on the surface defined by the equations
of motion (\ref{RPTeq}),
the torsion of {\it the world trajectory}
is reduced to the
constant,
$\varrho=-m\alpha^{-1}$,
whereas its curvature, $k$,
$k^2=x''_\mu x''{}^\mu$,
is conserved, $k^2=\alpha^2(m^2+p^2)$.
Identifying in (\ref{RPTeq})  the evolution parameter
with $x^0$ (laboratory time gauge), $x^0=\tau$,
and taking a non-relativistic limit,
$\vert \dot{\vx}\vert<<1$, we immediately find that
the equations (\ref{RPTeq}) with $\mu=i$ are reduced
exactly to (\ref{pxy}) when  $-\alpha$ is
identified with
the ``exotic" parameter $\kappa$.

To find the non-relativistic Lagrangian
of the model (\ref{Ptor}),
let us turn to the canonical formalism.
In accordance with
(\ref{RPTeq}), the
system (\ref{Ptor}) is described by the constraints
\begin{equation}
\chi\equiv p_\mu J^\mu+\alpha m=0,\qquad \psi\equiv P_q=0,
\label{Maj}
\end{equation}
responsible for reparametrization invariance.
The symplectic form and the Hamiltonian are
\begin{equation}
\begin{array}{ll}
\Omega_{tor}=dP_q\wedge dq+dp_\mu\wedge dx^\mu +
\frac{1}{2}\alpha^{-2}
\epsilon_{\mu\nu
\lambda}J^\mu
dJ^\nu\wedge dJ^\lambda,
\\[8pt]
H_{tor}=\alpha^{-1}q\chi +w\psi,
\end{array}
\label{tors}
\end{equation}
where
$w=w(\tau)$ is an arbitrary function \cite{MPtor}.
The $J_\mu$ take  values
in the two-sheeted hyperboloid, $J^2=-\alpha^2$,
and generate the $sl(2,R)$ algebra,
$\{J_\mu,J_\nu\}=-\epsilon_{\mu\nu\lambda}J^\lambda$.
The angular momentum
vector
\begin{equation}
\tilde{{\cal J}}_\mu=-\epsilon_{\mu\nu\lambda}x^\nu
p^\lambda + J_\mu
\label{Jang}
\end{equation}
is an integral of the motion
of the  system (\ref{Ptor}) and forms,
together with the energy-momentum vector
$p_\mu$, the Poincar\'e algebra
$iso(2,1)$.
The  first constraint
from (\ref{Maj}), being
the classical analog of the quantum (2+1)-dimensional
Majorana equation,
has three types of solutions mentioned above.
In the massive ($p^2<0$)
and the tachyonic ($p^2>0$) sectors
it defines the mass-spin relation,
where the
spin (central Casimir element) of $iso(2,1)$
is
${\cal S}=p\tilde{{\cal J}}/\sqrt{\vert p^2\vert}$.
In the Hamiltonian picture,
the gauge freedom is fixed
by imposing two gauge conditions
$x^0-\tau=0$ and
$q-\alpha/J^0=0$. Let us focus our attention
to the massive sector $p^2<0$,
and, again, consider the non-relativistic
limit $\vert \dot{\vx}\vert=\vert \vJ/J^0\vert <<1$.
Then the symplectic form $\Omega_{tor}$ reduces
to $\omega_{t}=dp_i\wedge dx_i+\frac{1}{2}\alpha^{-1}
\epsilon_{ij}dJ_i\wedge dJ_j$.
Using the Majorana-type constraint (\ref{Maj}),
the Hamiltonian becomes, in turn,
$H_t=p^0=\alpha^{-1}\vp\cdot\vJ-\frac{1}{2}m\alpha^{-2}
\vJ^2+m.$
Here we have taken into account the explicit dependence of
the
gauge condition $x^0-\tau=0$ on the evolution parameter,
and used that the positive energy sector $p^0>0$
corresponds to the upper sheet of the
hyperboloid with $J^0=\sqrt{\vJ^2+\alpha^2}$.
Identifying $\alpha^{-1}\vJ$
with $\vy$, and as before,
the $-\alpha$ with the ``exotic'' parameter $\kappa$,
and using the relations (\ref{LSZgoodcoordinates}),
we find that $\omega_t$
is reduced exactly to the symplectic
form (\ref{LSZsympham}),
whereas the Hamiltonian takes the form of the
Hamiltonian in (\ref{LSZsympham}) shifted by a
constant,
$H_t=\vp^2/2m -(2m\theta^2)^{-1}\vQ^2+m$.
This means that the non-relativistic
limit of the model (\ref{Ptor})
is described effectively by the Lagrangian
$L_t=L_I-m$, where $L_I$ is the
first order Lagrangian (\ref{FaJalag}).

In the non-relativistic limit,
the $iso(2,1)$ generators (\ref{Jang})
are reduced to
$\tilde{{\cal J}}_0={\cal J}+\theta m^2$,
$\tilde{{\cal J}}_j=
\varepsilon_{ij}{\cal K}_j$,
where ${\cal J}$ and ${\cal K}_i$
are integrals of motion
(\ref{LSZconsquant}),
whereas the $iso(2,1)$ spin ${\cal S}$ is reduced
to $\vQ^2/(2\theta)+\theta m^2$.
Finally, we note that the non-relativistic
limit of squared curvature,
$k^2=(x'')^2$, is $\vQ^2/(m\theta)^4$.
\vskip 0.3cm

In conclusion, we found that
   the model of Lukierski et al. \cite{LSZ}
can be decomposed into a union of slightly deformed
copies of the model
studied in \cite{Duval, Grig, DH}, which correspond to
non-relativistic anyons.
Our coordinates $\vX$ transform as vectors under Galilei
boosts, just like their non-commuting relativistic
counterparts for the anyon considered in
\cite{CortPlyush}. They are analogues of
the Foldy-Wouthuysen coordinates for the Dirac particle.
The commuting coordinates $\tilde{\vX}$
are in turn analogous to those localizable relativistic
coordinates.
The latter are,
however, not a Lorentz vector
\cite{CortPlyush},
and correspond rather to the
Newton-Wigner coordinates \cite{NW}.

Let us stress that the coordinates $X_i$ describing the
noncommutative plane have been identified by us
uniquely as the coordinates which form a
space vector
with respect to the transformations associated
with the exotic Galilei group generators
$p_i$, ${\cal J}$ and
${\cal K}_i$ from (\ref{LSZconsquant}).
They are translated by the Hamiltonian
(\ref{HJPK}) in contrast with
the initial coordinates
$x_i$ subjected to a Zitterbewegung.

The additional internal
degree of freedom,  $\vQ$, can be related as spin, since
the conserved angular momentum, ${\cal J}$ in
(\ref{LSZconsquant}),
reduces, in the rest frame, to $\vQ^2/2\theta$.
Remember that $\vQ$ is unaffected by Galilean boosts and
space translations.
In our view the proper motion of $\vQ$ is a gauge artifact;
the only physical quantity is $\vQ^2/2\theta$, which
is a constant of the motion, and
we identify
it with the anyonic spin.

Eliminating the negative--energy quantum
excitations amounts, furthermore, to suppressing
(reducing to a fixed value)
this internal spin degree of freedom and yields
   the minimal system studied in \cite{Duval, Grig, DH}.
The word ``minimal'' refers here to the action of the
(extended)
Galilei group, which is transitive in \cite{DH}, but not for
the case of the model \cite{LSZ}.
In a quantum language,
transitivity
corresponds to having an irreducible representation.
The elimination of the negative energy states, is,
however, {\it not} mandatory for a non-relativistic system.
(Just think of the Kepler problem, where the bound  motions
have negative energy).

We have observed that the interaction described
in terms of the covariant coordinates
$X_i$ results, at the quantum level,
in the Moyal product term
(\ref{Moyal}) in the corresponding Schr\"odinger equation.
We have also pointed out that while the exotic model
(\ref{exosymp}), (\ref{EXconsquant})
can be easily coupled
to an electromagnetic field,
the higher-derivative model (\ref{LSZlag})
does not
seem appropriate to accommodate a magnetic field
in a natural way.

We have demonstrated that the Lukierski et al.
system is the non-relativistic limit of the model of the
relativistic particle with torsion \cite{MPtor}.
In this limit the quantity $\vQ^2/2\theta$
is identified as a (shifted by a constant)
analog of the $iso(2,1)$ spin of the model (\ref{Ptor}),
and, on the other hand, is a rescaled
(by a constant) squared curvature
of the world trajectory of the particle (\ref{Ptor}).
As it was shown earlier \cite{MPtor,MPany, MPJN},
the reduction of the
model (\ref{Ptor}) to the surface of the constant
spin value ${\cal S}=-\alpha$ produces relativistic anyons,
whose non-relativistic limit is, in turn, the model of Duval
et al.,
cf. \cite{JNnon}.
This  is in a complete agreement with the relationship of
the models (\ref{LSZlag}),
(\ref{exosymp})-(\ref{EXconsquant}) and (\ref{Ptor})
elaborated here.

Having established the relation of the
non-relativistic
model \cite{LSZ} with its relativistic
counterpart \cite{MPtor},
we note that
the association of the indefinite metric
with higher-derivative term,
as it was done in
one of the alternative  quantization schemes
in \cite{LSZ}, does not seem
natural.
Indeed,
analogously to the relativistic case,
the indefinite metric prescription
is not consistent with the real nature
of the classical spin variables \cite{MPtor}.
On the other hand, it is known \cite{M0}
that the presence of higher derivatives
in Lagrangian in relativistic case
does not obligatorily leads to the appearance
of the tachyonic states in the spectrum.

At last, we note that the problem of introducing
interaction discussed here in the context of
non-relativistic anyons could give a new perspective
for analogous problem in the relativistic case \cite{CNP}.
The obtained results could also be helpful
for clarifying the question of
spin-statistics relation,
which still remains open
in the group-theoretical
approach to anyons \cite{CortPlyush}.

\vskip 0.5cm
{\bf Acknowledgements}.
A useful communication with J. Lukierski is gratefully
acknowledged.
One of us (PAH) is indebted to the
{\it Departamento de F\'{\i}sica} of {\it
Universidad de Santiago de Chile} for hospitality
extended to him.
The work was supported by the grants 1010073
and 7010073 from FONDECYT (Chile) and by DYCIT
(USACH).

\end{document}